\documentclass[10pt,twocolumn,superscriptaddress,amssymb,amsmath,aps,pra]{revtex4-2}
\usepackage{graphicx}
\usepackage[normalem]{ulem}
\usepackage{color}
\usepackage{graphicx}
\usepackage[normalem]{ulem}
\usepackage{color}

\usepackage{xpatch}
\xpatchcmd\bibsection{19}{5}{}{}
\xpatchcmd\bibsection{\begingroup}{\vskip -10pt\begingroup}{}{}

\begin{document}
\title{Quantitative in situ measurement of optical force along a strand of cleaved silica optical fiber induced by the light guided therewithin}
\date{September 16, 2021}
\author{Mikko Partanen}
\affiliation{Photonic Device Physics Laboratory, Department of Physics, 
Yonsei University, 50 Yonsei-ro Seodaemun-gu, Seoul 03722, Korea}
\affiliation{Photonics Group, Department of Electronics and Nanoengineering,
Aalto University, P.O. Box 13500, 00076 Aalto, Finland}
\author{Hyeonwoo Lee}
\affiliation{Photonic Device Physics Laboratory, Department of Physics, 
Yonsei University, 50 Yonsei-ro Seodaemun-gu, Seoul 03722, Korea}
\author{Kyunghwan Oh}
\affiliation{Photonic Device Physics Laboratory, Department of Physics, 
Yonsei University, 50 Yonsei-ro Seodaemun-gu, Seoul 03722, Korea}

\begin{abstract}
We proposed an optomechanical system to quantify the net force on a strand of cleaved silica optical fiber in situ as the laser light was being guided through it. Four strands of the fiber were bond to both sides of a macroscopic oscillator, whose movements were accurately monitored by a Michelson interferometer. The laser light was propagating with variable optical powers and frequency modulations. Experimentally, we discovered that the driving force for the oscillator consisted of not only the optical force of the light exiting from the cleaved facets but also the tension along the fiber induced by the light guided therewithin. The net driving force was determined only by the optical power, refractive index of the fiber, and the speed of light, which pinpoints its fundamental origin.
\end{abstract}

\maketitle

\section{Introduction}

In optomechanics, the motion of mechanical resonators is typically controlled by using laser light reflected from a mirror \cite{Gigan2006,Kleckner2006,Weld2006,Ma2015,Evans2014,Wagner2018,Wilkinson2013}. Therefore, the optical force, which drives these mechanical resonators is the well-known radiation pressure of light incident from free space. However, light can also exert forces while it is propagating inside materials, and in particular, when it is crossing material interfaces. Mostly, these forms of optical forces are studied in liquids by observing the deformation or movement of liquid surfaces under optical excitation \cite{Astrath2014,Ashkin1973,Casner2001,Choi2017,Schaberle2019,Zhang2015}. There also exist measurements of the steady-state radiation pressure on mirrors immersed in liquids \cite{Jones1954,Jones1978}. Experiments on the forces of light in lossless solids have been very scarce and rather qualitative or hindered by the lossy nature of solids \cite{Kundu2017,Gibson1970,She2008,Brevik2018b,Partanen2019e,Brevik2019,Brevik2009,Mansuripur2009b}. From literature, one can conclude that previous works have used neither mechanical resonators nor optical fibers to quantitatively measure optical forces inside materials. This is surprising and calls for a change since optical forces have an important role, e.g., in whispering gallery mode resonators \cite{Schliesser2008,Brevik2010} and as a source of cross-talk in multi-core fibers \cite{Diamandi2017}.

\begin{figure*}
\centering
\includegraphics[width=\textwidth]{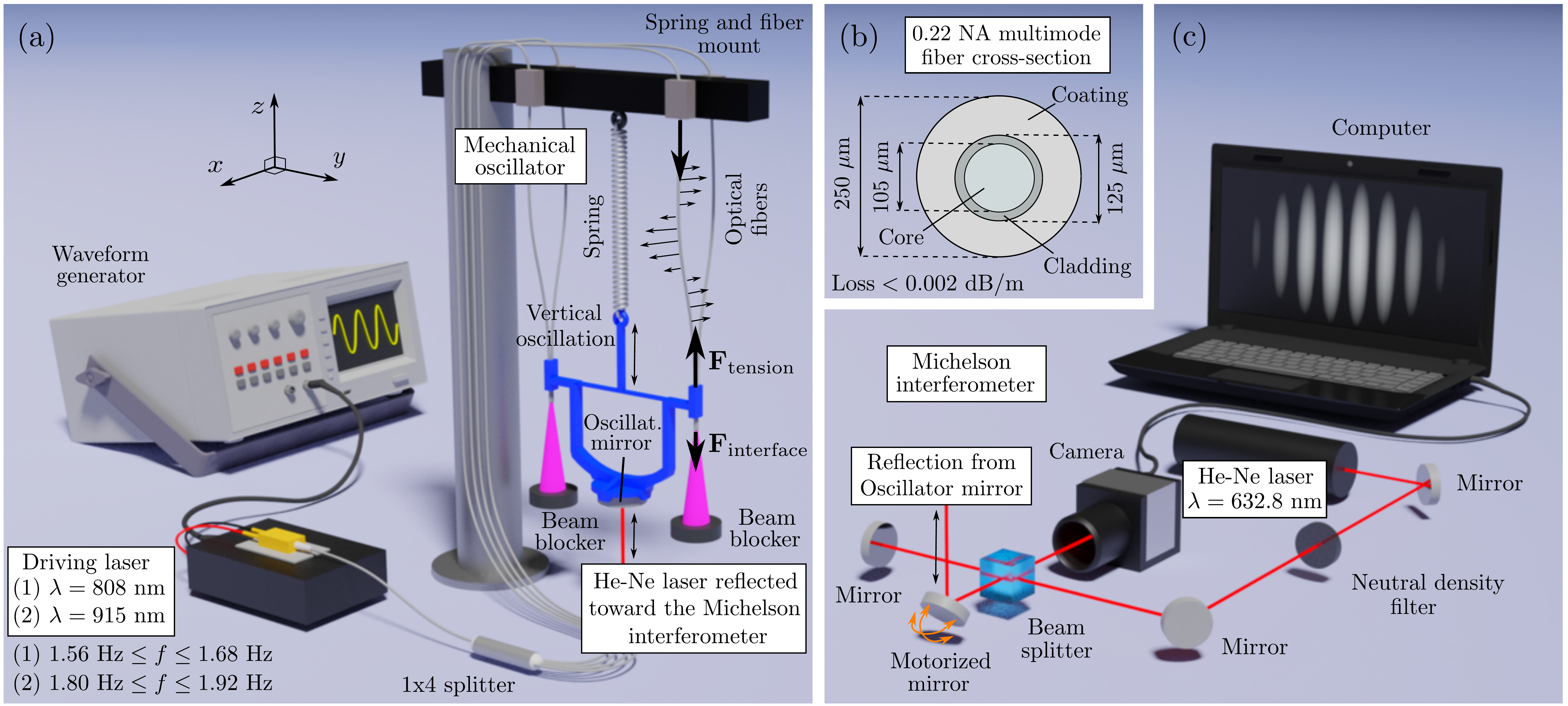}
\caption{\label{fig:setup} (a) The mechanical oscillator was driven by optical interface forces $\mathbf{F}_\mathrm{interface}$ and tension $\mathbf{F}_\mathrm{tension}$ of the four fibers, where the laser was propagating. These forces, illustrated for one of the fibers, were modulated by varying the laser intensity. The laser generated by a multimode laser diode was split into the four fibers by a 1x4 splitter and finally the laser exited from the ends of the four fibers, which were all bonded to the oscillator. The wavelength was either at 808 nm or 915 nm. (b) The cross-section of the 0.22 NA multimode fiber (Thorlabs, FG105LCA). (c) The nanoscale oscillation was detected by the Michelson interferometer utilizing a separate He-Ne laser. See the appendix for a more detailed description of the experimental setup.}
\end{figure*}

Among solid dielectric media, silica optical fiber is considered to be de facto the lowest-loss medium with the highest uniformity along its length \cite{Oh2012}, yet the forces of light carried by the optical fiber have not been quantified in experiments yet. In this work, we proposed a macroscopic oscillator platform to interferometrically quantify the forces of light, which is being guided through a silica optical fiber, for the first time to the best knowledge of the authors.  We used cleaved optical fiber strands bond to a macroscopic oscillator to make the light propagating through the optical fiber the only driving force of the oscillator. Our platform is schematically illustrated in Fig.~\ref{fig:setup}. A 3D printed mass was suspended by a spring at the center and, on both sides, the cleaved optical fibers were glued. Our experiment leaves no room for different interpretations of the origin of the oscillator signal but the forces of light propagating through the optical fibers. In particular, heating effects, whose consequences in many prior cases have dominated over the optical momentum transfer, are negligible since light propagating inside a commercial multimode optical fiber less than a meter experiences a loss less than 0.002 dB. Thermal effects are also hindered by the large thermal time constant of the macroscopic oscillator \cite{Ma2015,Partanen2020b}.

A 3D printed mass was designed to be held by a spring at the center and balanced by optical fiber strands glued at both ends as illustrated in Fig.~\ref{fig:setup}(a). A laser was split into four strands of commercial 0.22 NA multimode optical fibers (Thorlabs, FG105LCA) with equal optical power using a 1x4 splitter. The optical fibers, with cross-section in Fig.~\ref{fig:setup}(b), were cleaved at 90 deg and aligned vertically downward. To confirm the universality of our results, we studied two oscillators with different masses and damping constants, one driven at the wavelength of 808 nm and the other at 915 nm. The longitudinal displacements of the oscillators were detected by a Michelson interferometer in Fig.~\ref{fig:setup}(c) using a He-Ne laser and a mirror attached at the bottom of the mass. Shifts of the interference fringes were recorded with a CMOS camera at a frame rate of 200 frames per second for various incident laser powers and modulation frequencies. A more complete description of the experiment is presented in the appendix.

\section{Optical forces}

The optical forces induced by the light guided in the optical fiber can be divided into 
the force at the air-silica interface of the optical fiber facet, and the tension along the fiber strands as schematically shown by $\mathbf{F}_\mathrm{interface}$ and $\mathbf{F}_\mathrm{tension}$ in Fig.~\ref{fig:setup}(a). The force $\mathbf{F}$ on an object is equal to the temporal change of its momentum $\mathbf{p}$ as $\mathbf{F}=d\mathbf{p}/dt$. We study the momentum transfer of light at the end facet of optical fibers, where the light experiences the air-fiber core interface. The momentum of light in the optical fiber and the air is denoted by $p$, and $p_0$, respectively. Then, by summing over the four fibers of Fig.~\ref{fig:setup}(a), we obtain the magnitude of the total interface force from the conservation law of momentum as
\begin{equation}
F_\mathrm{interface}=\Big[T-\frac{p}{p_0}(1+R)\Big]\frac{P}{c}.
 \label{eq:radiationpressureforce}
\end{equation}
Positive and negative values of $F_\mathrm{interface}$ indicate whether the force is upward or downward in Fig.~\ref{fig:setup}(a), respectively, and $P=\sum_{i=1}^4P_i$ is the sum of the incident optical 
powers of the four fibers. The power reflection coefficient $R$, and the power transmission coefficient $T=1-R$ are equal in the four fibers. For the cleaved fiber facet without any thin-film coating, $R$ and $T$ are given by Fresnel formulas as $R=[(n-1)/(n+1)]^2$ and $T=4n/(n+1)^2$. Here $n$ is the refractive index of the fiber core and we set the refractive index of air to unity.

When light propagates along a bent fiber, the optical force pushes the fiber walls unequally at different sides and the net momentum flux of light changes its direction. This is a direct consequence of the momentum conservation law. The optical force on the fiber walls points in the direction of the positive curvature, which is always normal to the fiber as indicated by the small arrows in Fig.~\ref{fig:setup}(a). From the elasticity theory (see the appendix), it follows that this normal force gives rise to tension in the longitudinal direction of the fiber.
The net tension of the four fibers is given by
\begin{equation}
F_\mathrm{tension}=\sum_{i=1}^4\int_{A_i}({T_{zz,i}}-{T_{zz,i}^\mathrm{(0)}})dxdy=\frac{p}{p_0}(1+R)\frac
{P}{c}.
 \label{eq:tensionforce}
\end{equation}
Here $T_{zz,i}$ is the diagonal component of the stress tensor of the fiber $i$ in the vertical direction when the light is guided through of the fiber, $T_{zz,i}^{(0)}$ is the stress tensor component in the absence of the light, and $A_i$ is the total cross-sectional area of the fiber $i$ on the side of the oscillator.

From the optical interface force in Eq.~\eqref{eq:radiationpressureforce} and the tension along fibers in Eq.~\eqref{eq:tensionforce}, the net time-dependent driving force of the mechanical oscillator is given by
\begin{equation}
F=F_\mathrm{interface}+F_\mathrm{tension}=\frac{TP}{c}.
 \label{eq:netforce}
\end{equation}
The net force in Eq.~\eqref{eq:netforce} is interestingly independent of the value of the momentum of light inside the fiber since the dependence on $p$ is canceled out. It is, however, seen that this net force depends only on the optical power, refractive index of the fiber, and the speed of light, which pinpoints its fundamental origin.

\section{Mechanical oscillator}

For the mechanical oscillator used to detect the net laser-induced 
force, we write Newton's equation of motion as
\cite{Aspelmeyer2014}
\begin{equation}
\frac{d^2z}{dt^2}+2\zeta\omega_0\frac{dz}{dt}+\omega_0^2z=\frac{F}{
m},
 \label{eq:oscillator}
\end{equation}
where $m$ is the effective mass of the oscillator, $\omega_0$ is the undamped
resonance frequency, $\zeta$ is the damping 
coefficient, and $F$ is the net
driving force in Eq.~\eqref{eq:netforce}. The mechanical Q factor is defined in terms of the damping 
coefficient as 
$Q=1/(2\zeta)$. As the mass of the vertically aligned spring is 
not 
negligible, the 
effective mass of the oscillator is given by Rayleigh's value
$m=m_0+m_\mathrm{s}/3$, where $m_0$ is the mass of the oscillator and 
$m_\mathrm{s}$ is the mass of the spring \cite{Thomson1998}.

The net force due to a laser beam harmonically modulated with angular 
frequency $\omega$ is denoted by 
$F=F_0\cos^2(\frac{1}{2}\omega 
t)=\frac{1}{2}F_0[1+\cos(\omega t)]$, where $F_0$ is the peak to peak 
force amplitude. The steady-state solution of 
Eq.~\eqref{eq:oscillator} is given by
$z(t)=z(\omega)\cos(\omega t+\varphi)+F_0/(2m\omega_0^2)$, where
the displacement amplitude is
\begin{equation} 
z(\omega)=\frac{F_0/m}{2\sqrt{
(2\omega\omega_0\zeta)^2+(\omega^2-\omega_0^2)^2}}
\label{eq:displacement}
\end{equation}
and $\varphi=\arctan[2\omega\omega_0\zeta/(\omega^2-\omega_0^2)]\in[-\pi,0]$.
The resonance frequency of the underdamped oscillator with 
$\zeta<1/\sqrt{2}$ is $\omega_\mathrm{r}=\omega_0\sqrt{1-2\zeta^2}$. At $\omega_\mathrm{r}$, the displacement amplitude of the oscillator in 
Eq.~\eqref{eq:displacement} 
obtains its peak value, $z_0=F_0/(4m\omega_0^2\zeta\sqrt{1-\zeta^2})$. By measuring the
peak value of the displacement amplitude, one then obtains the driving force 
amplitude
as
\begin{equation}
 F_0=4m\omega_0^2\zeta\sqrt{1-\zeta^2}\,z_0.
 \label{eq:forceamplitude}
\end{equation}
Here $\omega_0$ and $\zeta$ can be 
accurately determined from the position and width of the mechanical 
resonance peak, and $m$ can be 
obtained from the oscillator, spring, and fiber masses measured with a 
digital scale \cite{Partanen2020b}. For our heavier oscillator, driven with the 808 nm laser, $m=(18.363\pm 0.001)$ g, while for the lighter oscillator, driven with the 915 nm laser, $m=(16.485\pm 0.001)$ g.

\section{Results and discussion}

The refractive index of the pure silica core of the fiber is $n=1.453$ at 
808 nm and $n=1.452$ at 915 m \cite{Malitson1965}. Then, at normal incidence, the 
transmission coefficient for the end interface of the fiber is 
$T=0.9659$ at 808 nm and $T=0.9660$ at 915 nm. For the 0.22 NA multimode fiber, 
the maximum angle of incidence in the fiber is near 8.7 deg for both 
wavelengths. The non-normal angles of incidence are accounted for in the 
analysis, even though their effect is small.

\begin{figure*}
\centering
\includegraphics[width=0.85\textwidth]{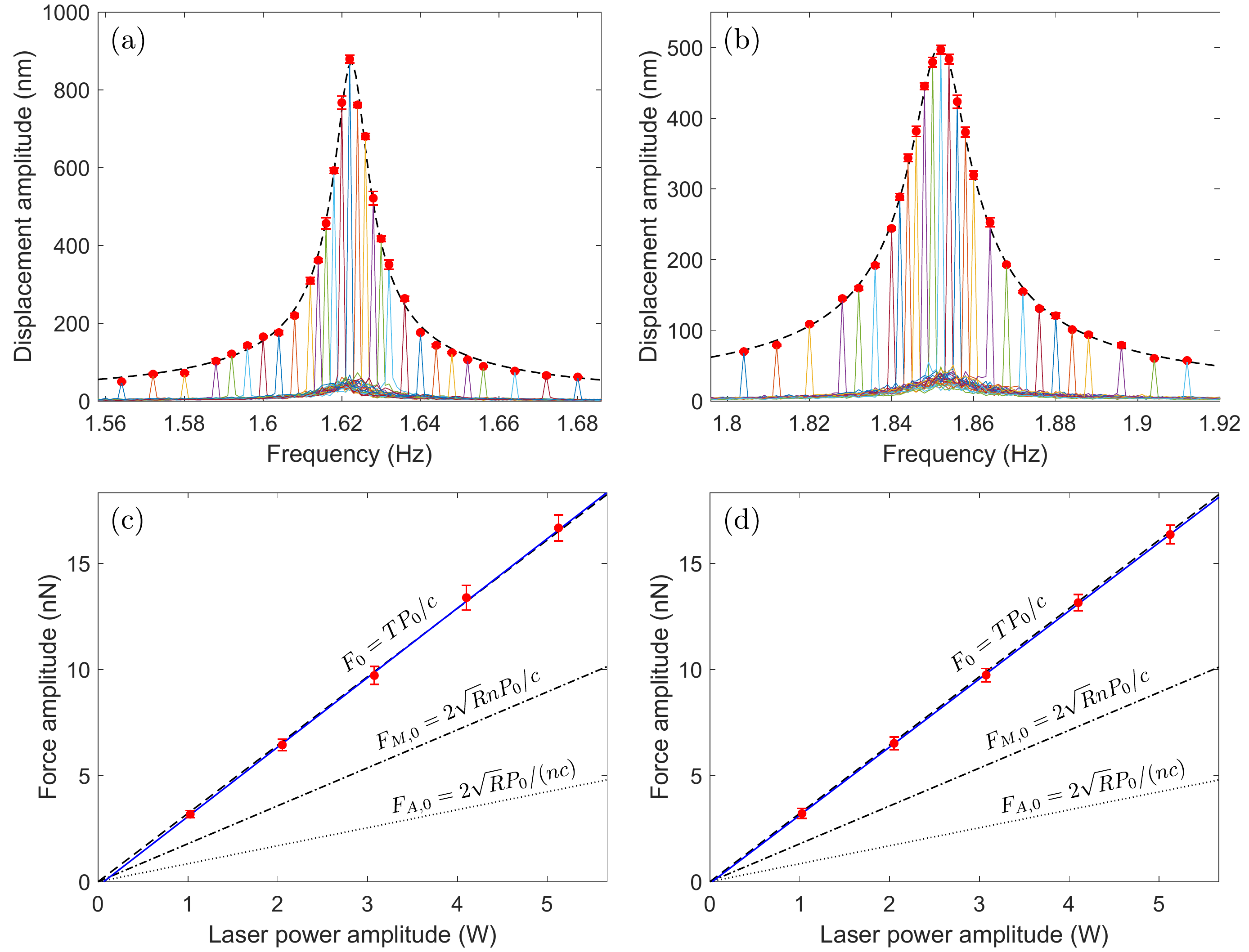}
\caption{\label{fig:results}
The displacement amplitude of the mechanical oscillator is plotted as a function of the laser modulation frequency (a) for the first oscillator at the wavelength of 808 nm and (b) for the second oscillator at the wavelength of 915 nm. The net peak-to-peak power amplitude of the driving field in the four fibers together is $P_0=5.125$ W in both cases. The solid line represents the averaged frequency spectrum measured with a single modulation frequency. The peak at each modulation frequency is marked with a red dot. The oscillator response function is fitted and shown by the dashed line. The corresponding peak-to-peak force amplitudes of the two oscillators are plotted in (c) and (d) as a function of the peak-to-peak laser power amplitude of the fibers. The solid lines represent the regression lines and the dashed lines show the net theoretical force in Eq. \eqref{eq:netforce}. The dash-dotted and dotted lines are the results of the Minkowski ($F_\mathrm{M,0}$) and Abraham ($F_\mathrm{A,0}$) momentum models, respectively, using Eq.~\eqref{eq:radiationpressureforce} with the corresponding momentum of light and excluding the tension in Eq.~\eqref{eq:tensionforce}.}
\end{figure*}

Figure \ref{fig:results} presents the experimental results. In 
Fig.~\ref{fig:results}(a), the measured displacement amplitude of the first 
oscillator driven at 808 nm is plotted as a function of the modulation 
frequency of 
the driving laser field. Fig.~\ref{fig:results}(b) presents the same plot for 
the second oscillator driven at 915 nm. The net peak to peak 
power amplitude of the four fibers used in these plots is 
$P_0=5.125$ W. The measurement time is an integer multiple of the modulation period 
close to 1000 s and the ensemble averaging is made over 10 or more 
measurements. As the error of the displacement amplitude, we use the standard 
deviation 
of the ensemble average.

In Figs.~\ref{fig:results}(a) and \ref{fig:results}(b), one can see that 
the 
fitted harmonic oscillator response function of Eq.~\eqref{eq:displacement} 
accurately describes the experimental results of both oscillators. One can also observe the
mechanical resonance peak in the noise spectrum that appears below 
the fitted response function. In the 
presence of photothermal effects, the response function 
would be modified from the ideal harmonic oscillator form as described, e.g., 
in Refs.~\cite{Ma2015,Ma2018}. In accordance with the results for the 
free space laser driven oscillator in Ref.~\cite{Partanen2020b}, the
photothermal effects are determined to be negligible for our macroscopic 
oscillator.

The fitting of the harmonic oscillator response 
function in the experimental data of the first oscillator in 
Fig.~\ref{fig:results}(a) gives the undamped 
frequency of the oscillator equal to
$f_0=(1.622411\pm 0.000088)$ Hz. The damping constant and the
Q-factor are found to be 
$\zeta=0.002485\pm 0.000063$ and $Q=201.2\pm 5.1$. The errors indicate the 
68.27\% confidence intervals of the fitting process corresponding to one standard deviation 
of normally distributed quantities.
The corresponding fitting using the experimental 
data of the second oscillator in 
Fig.~\ref{fig:results}(b) gives the 
undamped oscillator
frequency equal to $f_0=(1.851847\pm 0.000075)$ Hz and the damping constant 
equal to 
$\zeta=0.003690\pm 0.000051$, which corresponds to $Q=135.5\pm 1.9$.

Figures \ref{fig:results}(c) and \ref{fig:results}(d) present the measured peak 
to peak force amplitudes of the two oscillators 
following from Eq.~\eqref{eq:forceamplitude} as 
a function of the peak to peak laser power amplitude. The slope of the 
regression line of the first oscillator is 
$dF_0/dP_0=(3.28\pm0.10)\times10^{-9}\text{ s/m}=(0.982\pm0.034)/c$. 
The relative error is 3.5\%, from which 2.1\% comes from the determination of 
the damping constant and 1.4\% from the peak 
displacement amplitude. For the second 
oscillator, $dF_0/dP_0=(3.21\pm0.10)\times10^{-9}\text{ s/m}=(0.963\pm0.030)/c$. The 
relative error is 3.2\%, 
from which 1.2\% comes from the determination of 
the damping constant and 2.0\% from the peak 
displacement amplitude. The slope of the 
theoretical line from Eq.~\eqref{eq:netforce} is $T/c=3.22\times10^{-9}\text{ s/m}=0.966/c$. Thus, the experimental 
results 
of both oscillators
agree with the theory within the experimental accuracy.
Figures \ref{fig:results}(c) and \ref{fig:results}(d) also show that the results cannot be explained solely in terms of the interface forces of the conventional Minkowski ($p=np_0$) and Abraham ($p=p_0/n$) momentum models with Eq.~\eqref{eq:radiationpressureforce} \cite{Abraham1909,Abraham1910,Minkowski1908,Leonhardt2006a,Pfeifer2007,Barnett2010b,Barnett2010a,Partanen2017c,Bliokh2017a,Bliokh2017b,Partanen2017e,Kemp2011,Partanen2019a,Partanen2019b,Leonhardt2014,Milonni2010,Brevik1979,Partanen2018b}. Thus, accounting for the tension in Eq.~\eqref{eq:tensionforce} is necessary, in which case both conventional models give the same net force. The appearance of more than one force component in experimental setups can partly explain how the Abraham-Minkowski controversy has continued to this day.

To the best knowledge of the authors, the only previous measurement of forces of light exiting a solid medium is She's \emph{et al.}~report \cite{She2008}. In their measurements, She \emph{et al.}~were able to detect the recoil of a thin silica filament at the end where the light exited, but the results were not quantitatively accurate and their interpretation has raised subsequent debates \cite{Brevik2009,Mansuripur2009b}. In constrast, the present work can explain the observations of She \emph{et al.}~independently of the Abraham and Minkowski momentum models by including the tension in the fiber induced by the light guided therewithin.  A small bending of the filament before its end due to any asymmetry causes the light to create tension, which results in the net pushing effect on the end facet of the filament.

For a more detailed investigation of the refractive index dependence of the total oscillator force, the present experiment could be carried out by using optical fibers with different refractive indices. The refractive index would need to deviate from that of fused silica by an amount that would change the experimental results sufficiently compared to the error bars of the experiment. Presently, there are no such solid-core optical fibers available that would also have low loss. The experiment could also be carried out by using a hollow-core photonic crystal fiber.

\section{Conclusion}

In summary, we have demonstrated that optical forces of light in an optical fiber can be quantitatively measured in situ while the fiber guides the light.  We proposed an optomechanical system where a macroscopic mechanical oscillator was driven only by the optical force in the fiber and its nanoscopic displacements were monitored interferometrically. The light guided along an optical fiber provided two forces, the interface force at the end facet and the tension on a curvature. The theoretical model agreed with the experimental measurements with an accuracy of 3.5\%.
Our work can pave the way to more extensive use of novel mechanical resonator geometries for accurately detecting optical forces in solids, e.g., by utilizing whispering gallery modes \cite{Brevik2010}.

\begin{acknowledgments}
This work has been funded by European Union's Horizon 2020 Marie 
Sk\l{}odowska-Curie Actions (MSCA) individual fellowship 
under Contract No.~846218 and the National Research Foundation of 
Korea (NRF) grant by the Korea government (MSIT) under Contract 
No.~2019R1A2C2011293.
\end{acknowledgments}

\appendix

\section{}

\subsection{Fabrication and design of the mechanical oscillator}

The mechanical oscillator masses were fabricated by 3D printing using the fused  deposition modeling (FDM) technique. The printing material was polylactic acid, commonly known as PLA. As illustrated in Fig.~\ref{fig:setup}(a), the designs of the oscillator masses included a mirror mount, two side arms to which the optical fibers were bonded, and a hook connecting the oscillator mass to the mechanical extension spring, which carried the weight of the oscillator mass. The total width of the oscillator between the attachment points of the optical fibers was about 9.0 cm. The rest masses of the first and second mechanical oscillators were 15.561 g and 13.683 g, respectively. These masses include 6.715 g of the mass of the mirror (Thorlabs, BB1-E02) that was mounted on the bottom of the oscillator.

\subsection{Driving laser and the optical fiber}

The driving laser beam at 808 nm was generated by a multimode laser diode module 
(Box Optronics, BLD-F808-06-22N0) and the laser beam at 915 nm was generated 
by a multimode laser diode module 
(Box Optronics, BLD-F915-10-22N0). The laser beams were coupled to a commercial 
low-loss 0.22 NA silica core 
multimode fiber (Thorlabs, 
FG105LCA). The core diameter of the fiber was $(105\pm2)$ $\mu$m, the cladding diameter was $(125\pm1)$ $\mu$m, and the coating diameter was $(250\pm10)$ $\mu$m.
The intensity of the driving 
laser beam was modulated by 
a waveform generator (Agilent, 33120A) connected to a laser driver 
(Arroyo Instruments, LaserPak 485-08-05). The temperature of the laser was 
controlled with a temperature controller (Arroyo Instruments, TECPak 
585-04-08). Before the mechanical oscillator, the laser beam was split into 
four beams by a fiber optic 1x4 splitter (Lfiber, IRBC-105-1-4-L1-N). Above the 
mechanical oscillator, the four fibers were mounted on an iron bar so that there 
was 3.0 mm loose in the fibers above the points where they were attached to the oscillator. This is described in more detail below.
Due to the smallness of the absorption coefficient of the optical fiber and the 
large
time 
constant of the macroscopic oscillator, photothermal effects had negligible 
influence on our 
experimental results.

\subsection{Measurement of the laser power}

The laser power 
was measured at the ends of the optical fibers by using an optical power meter 
(Thorlabs, PM400) with an integrating 
sphere sensor (Thorlabs, S145C). The 
nonzero reflectivity ($R=0.0341$ at 808 nm and $R=0.0340$ at 915 nm) and the corresponding nonunity 
transmissivity ($T=0.9659$ at 808 nm and $T=0.9660$ at 915 nm) of the fiber ends were accounted for in the analysis. The optical powers propagating through all four fibers after the fiber optic splitter were measured to be equal within 1.0 \%. Thus, the laser-induced forces on the two sides of the oscillator were well balanced.
The used peak to peak amplitude of the sinusoidal modulation was 1.0 \% smaller in 
comparison with the 
otherwise stationary laser beam. Thus, for example, for a stationary beam with 
a power of 5.000 W, after adding the modulation and accounting for the 
nonunity transmissivity of the fiber ends, the peak to peak power incident to 
the fiber ends becomes 5.125 W. This is the value used in the analysis corresponding to the results in Figs.~\ref{fig:results}(a) and \ref{fig:results}(b).

\subsection{Mechanical extension springs}

In the experimental setup, the mechanical oscillator was hanging on three 
hard extension springs used in series to 
obtain a relatively small total spring constant. The springs 
were made of music wire and 
they had cross-over type hooks at their ends. The upper and lower springs 
were of equal type (Acxess Spring, 
PE016-312-129000-MW-2500-MH-N-IN) having a mass of 
$m_\mathrm{s,1}=3.025$ 
g and
a reported extension rate of $k_\mathrm{s,1}=5$ N/m. The middle spring 
(Acxess Spring, 
PE016-312-90250-MW-1880-MH-N-IN) had a mass of $m_\mathrm{s,2}=2.221$ g and a 
reported extension rate of 
$k_\mathrm{s,2}=7$ N/m. The total mass of the four optical fibers parallel to the springs between the holder and the oscillator was 0.136 g, and the total mass of the springs and the fibers was equal to $m_\mathrm{s}=8.407$ g. Both 
the extension rates above and the masses of 
the vertically aligned springs and fibers contribute to the 
total spring constant of the oscillator. In the analysis of 
our experimental results, we 
used the effective mass, undamped angular frequency, and the damping constant as 
the only oscillator parameters. The total spring 
constant of the 
system can be determined from the experimental results as 
$k=m\omega_0^2$. This gives for the first oscillator the total spring constant 
of $k=1.908$ N/m, and for the second oscillator we obtain $k=2.232$ N/m.

\subsection{Acoustic and seismic isolation}

The experimental setup was mounted on an actively damped optical table for 
isolating 
it against acoustic and seismic vibrations. The mechanical oscillator 
part of the setup was also protected against air flows by covering it with plastic walls. To minimize any disturbances in the 
surroundings of the laboratory, the measurements were carried out at nighttime.

\subsection{Michelson interferometer}

The motion of the mechanical oscillator was detected 
by a Michelson interferometer. The interferometer was used to monitor the motion of the oscillator by setting the oscillator mirror in one of the 
two interferometer arms. The arm length of the 
interferometer was about 10 
cm. One of the interferometer arm mirrors was 
motorized and it could be used for tuning the interference 
fringe spacing remotely. However, the fringe spacing was adjusted only before each 
measurement and 
it was not actively changed during the experiments.
All mirrors in the interferometer arms (Thorlabs, BB1-E02) had a reflectivity 
of over 
99\%. The interferometer laser was a 5-mW continuous-wave 
TEM$_{00}$ He-Ne 
laser (JDSU, 1125P) operating at 632.8 nm. Before the beam splitter, the interferometer laser power was reduced by a factor of 1/10 
by a neutral density
filter (Thorlabs, NE10A). The motion of the interference fringes during the experiment was recorded 
by a CMOS camera (Edmund Optics, EO-0413C) that was connected to a computer. The 
frame size recorded was 600x30 pixels and the recording was made with a frame 
rate of 200 frames per second.

\subsection{Tracking the motion of interference fringes}

The horizontal movement of the interference fringes was tracked from the recorded video 
files by monitoring the positions of the intensity maxima and minima in each frame. The fringes are illustrated in the computer screen of 
Fig.~\ref{fig:setup}(c). When the 
fringes move a distance that is equal to the distance between two intensity 
maxima, the mechanical oscillator moves half a wavelength 
in 
the vertical direction.
For efficient analysis of millions of frames in total, we used a C++ code utilizing
the Open Computer Vision Library (Open CV). The analysis followed the same approach as in Ref.~\cite{Partanen2020b}. The mechanical oscillator, which was hanging on a spring, could move in all three dimensions, but the interferometer was the most 
sensitive for the vertical motion of the oscillator that was of our interest. 
If the oscillator was disturbed by a large amount, the scale of the interference fringes could vary, and the fringes could also rotate. These effects were, however, negligibly small when 
external noise sources were minimized 
during measurement conditions.

\begin{figure}[b]
\vspace{-0.3cm}
\centering
\includegraphics[width=0.9\columnwidth]{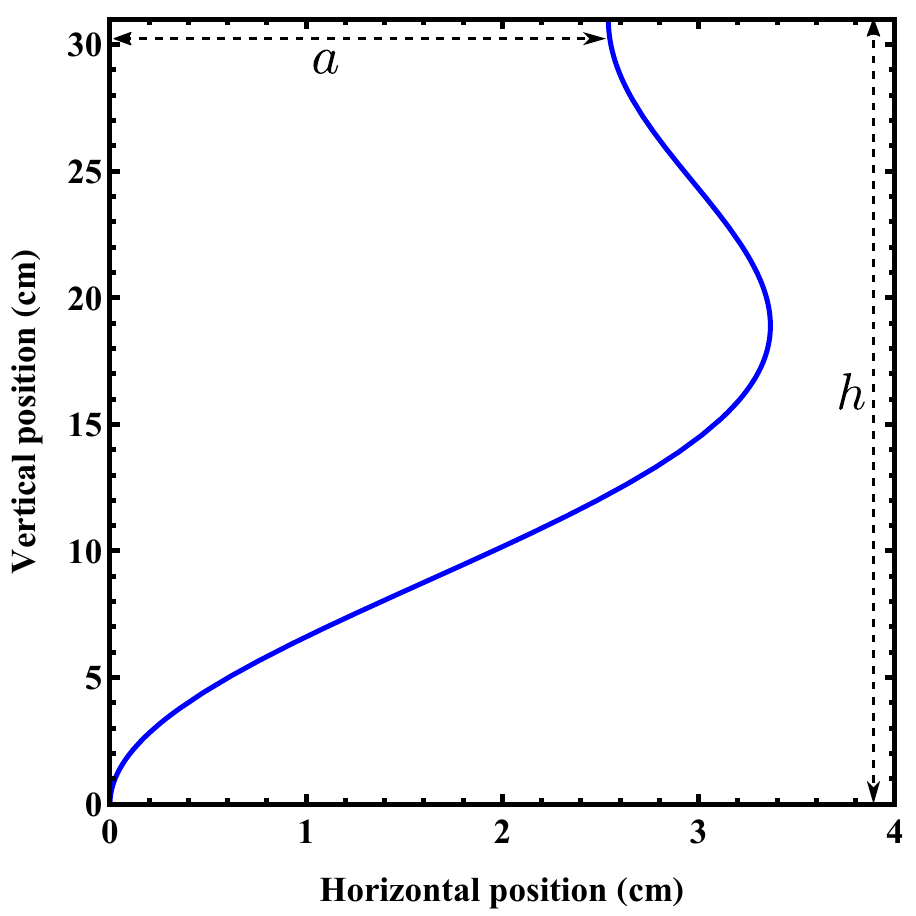}
\vspace{-0.1cm}
\caption{\label{fig:fiberpath}
Illustration of the fiber path and the related horizontal and vertical distance parameters $a$ and $h$.}
\end{figure}

\subsection{Calculation of the tension along the fiber}

This section describes how the elasticity theory calculations are used to verify the accuracy of the analytic result in Eq.~\eqref{eq:tensionforce}. The optical force on the fiber walls due to the small bending of the fiber follows directly from the conservation law of momentum. It depends on the local curvature of the fiber and can be used in the elasticity theory simulations of the behavior of the fiber. When $x=g(z)$ is the curve of the fiber, the optical force density due to the bending of the fiber averaged over the cross-sectional area of the fiber is equal to $\mathbf{f}(z)=(p/p_0)P/(cA)g''(z)/(g'(z)^2+1)^2(g'(z)\hat{\mathbf{z}}-\hat{\mathbf{x}})$, where $p$ is the momentum of light in the fiber, $p_0$ is the momentum of light in vacuum, $c$ is the speed of light in vacuum, $A$ is the cross-sectional area of the fiber, and $\hat{\mathbf{x}}$ and $\hat{\mathbf{z}}$ are unit vectors along the $x$ and $z$ axes, respectively. The optical power $P$ in the force density formula above accounts for both the incident and reflected powers. At the end points of the fiber, $z_1=0$ and $z_2=h$, the curve of the fiber satisfies $g'(z_1)=g'(z_2)=0$, which means that the ends of the fiber between the mechanical oscillator and the fiber mount above it are vertically aligned. We have parametrized the physical path of the fiber as $g(z)=\frac{a}{2s}\{[1+\sin(\frac{2\pi}{h}z-\frac{\pi}{2})]+s[1+\sin(\frac{\pi}{h}z-\frac{\pi}{2})]\}$, where $a$ and $h$ are physical distance parameters illustrated in Fig.~\ref{fig:fiberpath} and $s$ is a parameter that controls the amount of loose in the fiber. The length of the fiber is $L=\int_ 0^h\sqrt{g'(z)^2+1}\,dz$ and the loose in the fiber is $d=L-L_0$, where $L_0=\sqrt{a^2+h^2}$ is the direct distance between the fiber end points. The experimental horizontal and vertical distance parameters are $a=2.54$ cm, $h=31.0$ cm. The parameter $s$ is determined so that the loose in the fiber is $d=3.0$ mm, which gives $s=1.35$.

We used the force density above in the elasticity theory simulations performed by using Comsol Multiphysics simulation tool with realistic dimensions and material parameters of the silica fiber. The material parameters used in the simulations include the mass density of $\rho=2200$ kg/m$^3$ \cite{Bruckner1970}, Young's modulus of $Y=70$ GPa, Poisson's ratio of $\nu=0.15$ \cite{DeJong2000}, and the refractive indices of the pure silica core of $n=1.453$ at 808 nm and $n=1.452$ at 915 nm \cite{Malitson1965}. The power reflection and transmission coefficients corresponding to these refractive indices are $R=0.0341$ and $T=0.9659$ at 808 nm and $R=0.0340$ and $T=0.9660$ at 915 nm.

\begin{figure}
\centering
\includegraphics[width=0.9\columnwidth]{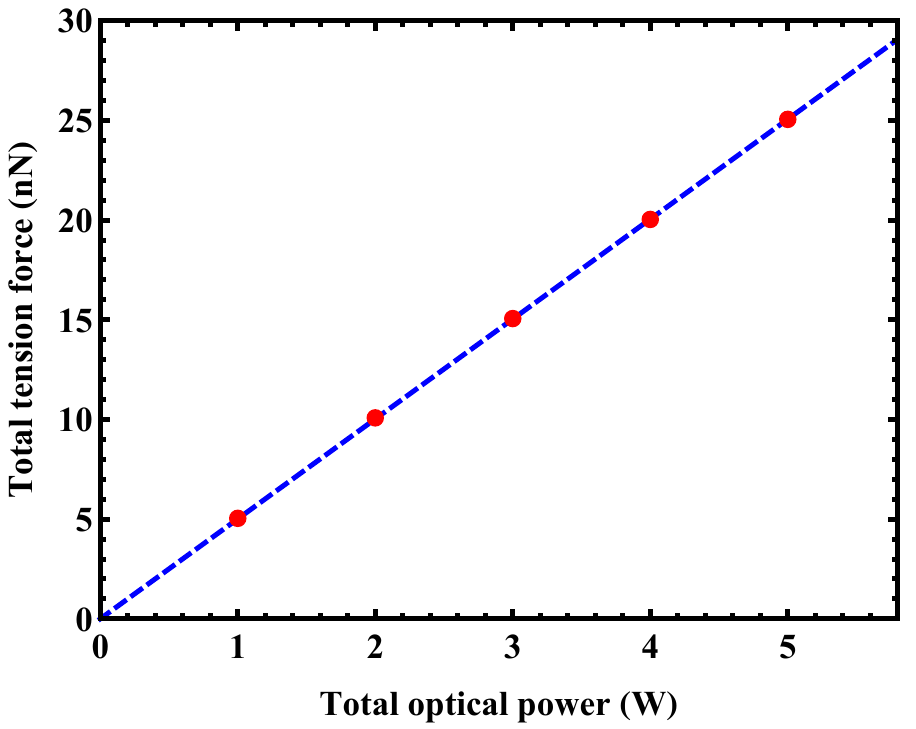}
\vspace{-0.1cm}
\caption{\label{fig:tensionforce}
Elasticity theory simulation of the total tension force of four fibers on the oscillator as a function of the total incident optical power propagating through the fibers. The dots represent the simulation results obtained by using the expression on the left hand side of Eq.~\eqref{eq:tensionforce} and the dashed line is the theoretical line, given on the right hand side of Eq.~\eqref{eq:tensionforce}.}
\vspace{-0.5cm}
\end{figure}

The simulation results of the total tension force of all four fibers above the oscillator as a function of the total incident optical power propagating through the fibers at the wavelength of 808 nm are illustrated in Fig.~\ref{fig:tensionforce}. The tension is calculated at the fiber end attached to the oscillator, i.e., at the origin of Fig.~\ref{fig:fiberpath}. In the simulations, we have used the Minkowski momentum in the optical force density with $p/p_0=n$,  but it only scales both the simulation result and the theoretical line, so the conclusions on the equality of Eq.~\eqref{eq:tensionforce} also apply to other values of $p/p_0$. It is seen that the simulation results hit the theoretical line on the right hand side of Eq.~\eqref{eq:tensionforce} within the 0.1\% numerical accuracy of the simulations. Thus, the tension force is accurately linearly proportional to the optical power in agreement with Eq.~\eqref{eq:tensionforce}. Equally accurate results are obtained for the second experimental wavelength of 915 nm, but they are not illustrated here. In the simulations, the steady state tension force for the constant optical power used in the simulations is obtained in a time scale that is short compared to the time scale of the harmonic modulation of the optical force. Therefore, the same linear proportionality between the optical\linebreak

\newpage

\noindent power and the tension force can be assumed for the harmonically modulated laser beam used in the experiment.

We have also made simulations for several values of loose in the fiber in the range $d<2$ cm. These calculations show that the resulting tension along the fiber is insensitive to small amounts of loose in the fiber provided that the ends of the fiber are vertically aligned as described above (i.e., the results correspond to those illustrated in Fig.~\ref{fig:tensionforce} within the 0.1\% numerical accuracy of the simulations). Therefore, the equality in Eq.~\eqref{eq:tensionforce} can be assumed to be accurate in the theoretical analysis of the results. Even though the tension force is independent of the amount of loose in the fiber, the loose in the fiber affects the time-dependent dynamics of the oscillator by modifying the damping constant and the resonance frequency of the oscillator. These are parameters that are determined experimentally as described above.


\end{document}